\documentstyle[twoside,fleqn,espcrc2]{article}

\newcommand{\be}{\begin{equation}}
\newcommand{\ee}{\end{equation}}
\newcommand{\bea}{\begin{eqnarray}}
\newcommand{\eea}{\end{eqnarray}}
\newcommand{\YM}{Yang-Mills }
\newcommand{\NC}{N}
\newcommand{\SUN}{SU(\NC)}

\newcommand{\SUT}{SU(2)}

\newcommand{\Real}{{\mathbf R}}
\newcommand{\Iden}{{\mathbf 1}}

\begin{document}

\title{Nahm transformation on the lattice\thanks{Poster presented by C. Pena at 
Lattice '99, Pisa, Italy.}
\vskip-3cm\hfill\small FTUAM-99-26; IFT-UAM/CSIC-99-30\vskip2.6cm
}

\author{A. Gonz\'alez-Arroyo\address{ Departamento de F\'{\i}sica Te\'orica C-XI,
        Universidad Aut\'onoma de Madrid, Madrid 28049, Spain.}
\address{ Instituto de F\'{\i}sica Te\'orica C-XVI,
        Universidad Aut\'onoma de Madrid, Madrid 28049, Spain.}
and C. Pena ${}^{\rm a}{}$}

\begin{abstract}
The Nahm transformation is a duality mapping between self-dual \YM configurations
on the torus, which exchanges the number of colours with the topological charge.
We show how lattice gauge theory techniques can be used to implement it
numerically. The method is presented and its precision illustrated with some applications.
\end{abstract}

\maketitle

\section{Introduction}

The Nahm transformation~\cite{Nahm,BVB} is a powerful mathematical tool to study self-dual
\YM fields living
on a 4-torus. It can furthermore be extended to the more general
case of fields on $T^{4-n} \times \Real^n,~n = 0,\ldots,4$
\footnote[1]{When noncompact dimensions happen
some modifications in the construction are required. However, as we will always work on $T^4$
this problem will not appear.}. In particular, when $n=4$ the ADHM
construction is recovered within this context. Some of the problems in which
the Nahm transformation has succeeded are the construction of instanton
solutions ({\em calorons}) with non-trivial holonomy at finite temperature ($n=3$)~\cite{caloron,calchi},
together with their associated fermion zero modes (ZM)~\cite{calZM},
or a nonexistence
proof for $Q=1$ fields with trivial twist on $T^4$~\cite{BVB}.

A limitation to the usefulness of the Nahm transformation comes from the lack of
analytic knowledge on self-dual fields on $T^4$; these can be,
on the other hand, accurately approximated and studied by means of lattice
techniques\footnote{See~\cite{tonyrev} for a recent review on both the analytical
and numerical sides.}.
It is therefore desirable to develop a numerical version of the Nahm
transformation such that it can be applied directly on lattice configurations.
Here we report on the lattice implementation of the Nahm
transformation we have set up, together with some examples of its abilities and
accuracy; some of the results obtained from its application are presented in~\cite{tonylat99}.

\section{The Nahm transformation}

Here we remind briefly some basics of the Nahm transformation.
Let us consider a $\SUN$ self-dual gauge field $A_{\mu}$ of topological charge $Q$
on a 4-torus, and construct a Weyl operator
$\bar{D}_z(x)=\bar{\sigma}_{\mu}(D^{\mu}(x)-2\pi i z^{\mu})$, with
$\bar{\sigma}_{\mu}=(\Iden,-i\vec{\tau})$ ($\tau_i$ being the Pauli matrices), and
$z^{\mu}\in \Real$.
Now, from the index theorem it follows that, provided $A_{\mu}$ contains no trivial flat
factors, the operator $\bar{D}_z$ has $Q$ ZM satisfying the Weyl equation $\bar{D}_z(x) \Psi^{\alpha}_z(x)=0$,
where $\alpha$ runs in the $Q$-dimensional space of ZM. The
{\em Nahm transform} of $A_{\mu}$ is defined as:
\be
\label{NahmDual}
\hat{A}_{\mu}^{\alpha \beta}(z) = i \int d^4x \ \Psi^{\alpha}_z(x)^{\dagger}
\frac{\partial}{\partial z_{\mu}} \Psi^{\beta}_z(x) \ \ ,
\ee
where a trace is taken over all other indices in $\Psi$.

The basic properties of the transformation are:
{\bf (1)}
$\hat{A}_{\mu}$ is a self-dual $SU(Q)$ potential with topological charge $\NC$, living
on the dual torus;
{\bf (2)}
applying it on $\hat{A}_{\mu}$ gives back $A_{\mu}$; and
{\bf (3)}
it induces an isometry between the original and dual moduli spaces.

The above construction works only in the case of trivial twist, i.e. when the
twist vectors $\vec{k},~\vec{m}$ vanish modulo $\NC$, due to the fact that
fermion fields transforming in the fundamental representation of $\SUN$ do
not support consistently nontrivial twisted boundary conditions. This can be circumvented,
however, in two equivalent ways~\cite{NahmTBC}: either by replicating
the twisted configuration along certain directions until an untwisted field is obtained,
or by introducing a new {\em flavour} index on which a twist is imposed that
compensates for the one acting over colour. The transformation can thus be applied on
{\em any} self-dual field on the torus\footnote{Both constructions have
been implemented within our numerical framework, and prove to be perfectly equivalent.
The results in Fig.~1 have been obtained
with replicas, the ones in Fig.~2 with flavoured fermions.}.
Properties {\bf (1)}-{\bf (3)} still hold, with $Q$ replaced by $N_0 Q$ and $\NC$ by
$\NC/N_0$, $N_0$ being a twist-dependent integer. In this general case,
the torus where $\hat{A}_{\mu}$ lives is not the dual one (see~\cite{NahmTBC} for details).

\section{Numerical method and sample results}

Our implementation of a lattice Nahm transformation involves two main stages:
first, we construct a lattice gauge configuration whose transform interests us;
for this we use improved cooling~\cite{impcool} techniques,
which provide us with a tight control over the structure of the field.
Then, the relevant quantities entering the transformation
itself, such as fermion ZM, must be computed. Here we describe
how to construct $\hat{F}_{\mu \nu}$, as well as ``dual'' link variables $\hat{U}_{\mu}$, the latter
allowing to calculate such quantities as Wilson loops.

For the computation of $\hat{F}_{\mu \nu}$ we use the identity:
\be
\label{fmunu} 
\hat{F}_{\mu\nu}(z) =  8\pi^2 \eta_{\mu \nu}^a \int d^4x  \Psi_z^{\dagger}(x) \tau_a
 \chi_z(x)
\ \ ,
\ee
where $\eta_{\mu \nu}^a \sigma_a = 
\bar{\sigma}_{\lbrack \mu} \sigma_{\nu \rbrack}$, and $\chi_z$ fulfills:
\be
\label{inh}
D_z^{\mu}(x)D_z^{\mu}(x) \chi_z^{\alpha}(x) = \Psi_z^{\alpha}(x) \ \ .
\ee

We substitute directly all these continuum expressions by their lattice versions.
Through the whole computation we use naive fermions; doublers will therefore appear, but
it is always possible to isolate the physical modes, as explained
below. Alternatively, Wilson fermions can be introduced from the beginning to spoil
doublers; this technique was in fact implemented first, and is described in detail in~\cite{nahm1}.

The first step is the computation of lattice ZM in the
background of a given self-dual field and for a given $z$.
As no exact ZM will appear,
what we do is to search for the lowest eigenvectors of the hermitian positive operator
$\bar{D}^{\dagger}_z(x)\bar{D}_z(x)$
by using a standard {\em conjugate gradient} algorithm, which supplies both high stability and
accurate solutions. Thus we end with
a lowest $8Q$-dimensional (for $\SUT$) eigenspace of quasi-ZM,
while higher modes carry ${\cal O}(1)$ eigenvalues.
The subspace of physical ZM is selected by diagonalising the Wilson-Dirac operator
within this $8Q$-dimensional lowest space. In this way one gets $Q$ exactly chiral
lowest modes; this is the multiplicity given by the index theorem.

Once the ZM have been computed, the operator inversion involved in Eq.~(\ref{inh})
must be carried out. For this we use a {\em stabilised biconjugate gradient}
algorithm~\cite{BiCGStab}, which proves highly efficient.
Finally, we compute $\hat{F}_{\mu \nu}$ in Eq.~(\ref{fmunu}).

\begin{figure}[htb]
\vspace{2.2cm}
\includegraphics{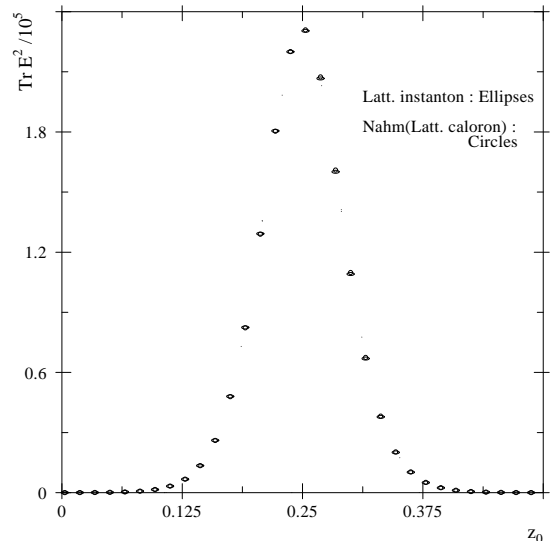}
\vspace{3.3cm}
\caption{Action density of a $\SUT$ $Q=1/2$ instanton-like configuration
with twist $\vec{k}=\vec{m}=(0,1,0)$
living on a $8^3 \times 32$ lattice vs. the Nahm transform of a $16^3 \times 4$ $\SUT$ $Q=1/2$ caloron-like
field with the same twist at coincident dual points.}
\end{figure}
\vspace{-0.75cm}

The whole computational procedure turns out to be much faster than the
one described in~\cite{nahm1}. Their results, on the other hand,
agree within a high precision; thus we have two alternative, mutually consistent frameworks.

A remarkable feature of our implementation is its accuracy, which can be tested in a number of
examples of known behaviour under Nahm duality. To illustrate this point we present the comparison in Fig.~1,
in which a highly precise matching is displayed between two action densities which
are predicted to be identical~\cite{dualities}.

It is possible to construct a {\em lattice Nahm transform}, which is useful, for instance, in computing
Wilson loops for the Nahm-dual field. The Nahm-dual lattice link variable $\hat{U}_{z,z+\Delta}$
is built up by first computing the lattice ZM $\Psi^{(\ell)}$ at $z$ and $z+\Delta$;
then the matrix ${\cal U}(z) = \sum_{x \in latt.} \Psi^{(\ell)}_z(x)^{\dagger} \Psi^{(\ell)}_{z+\Delta}(x)$
(which approximates a Wilson line, cf. Eq.~(\ref{NahmDual})) is formed; and, finally, it is decomposed
as ${\cal U}(z)= H(z) \hat{U}_{z,z+\Delta}$, with $H(z)$ a hermitian positive matrix. The link
is thus unitary and gauge-covariant by construction.
We exemplify the procedure by applying it to a vortex-like $\SUT$ configuration as those discussed in Ref.~\cite{vortex}, which
is Nahm-self-dual. The Wilson loop of radius $r$ centered around the vortex in the Nahm transform is shown in Fig.~2, and displays
the expected interpolation between 1 and -1.

\vspace{-3mm}
\begin{figure}[htb]
\vspace{1.5cm}
\includegraphics{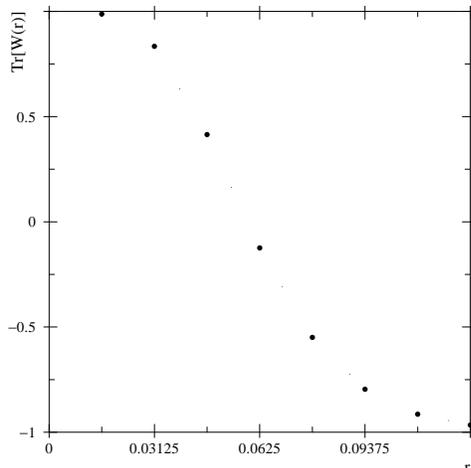}
\vspace{33mm}
\caption{Wilson loops revealing a vortex-like structure in the dual of a $Q=1/2$ lattice field,
with twist $\vec{k}=\vec{m}=(0,1,0)$, living in a $4 \times 16 \times 4 \times 16$ lattice.
The dual torus large periods have length $1/2$;
the loop radius $r$ ranges in $\lbrack 0,1/8 \rbrack$.}
\end{figure}
\vspace{-12mm}

\section{Conclusions}

\vspace{-1mm}

A lattice implementation of the Nahm transformation, useful to study the properties of
self-dual \YM fields, has been set up in a consistent and
perfectly satisfactory way. Several checks have been performed, and a remarkable
accuracy is featured by the results.

The method has already been successfully applied to discover new properties
of non-abelian, twisted gauge fields living on a torus~\cite{nahm1,dualities}. However,
its possibilities are far from being exhausted, and a number of interesting
applications are in sight.

\vspace{3mm}

{\bf Acknowledgements}: We thank Margarita Garc\'{\i}a P\'erez and
\'Alvaro Montero for many fruitful discussions.


\vspace{-2mm}

\begin{thebibliography}{9}

\bibitem{Nahm}
W.~Nahm, Phys. Lett. B90 (1980) 413; in
{\em Lecture Notes in Physics} 201, eds. G. Denardo e.a. (1984) p.189.

\bibitem{BVB}
P.~Braam and P.~van~Baal, Commun. Math. Phys. 122 (1989) 267.

\bibitem{caloron}
T.C. Kraan and P. van Baal, Phys. Lett. B428 (1998) 268; Nucl. Phys. B533 (1998) 627.

\bibitem{calchi}
K. Lee and P. Yi, Phys. Rev. D56 (1997) 3711; 
K. Lee and C. Lu, Phys. Rev. D58 (1998) 025011.

\bibitem{calZM}
M.Garc\'{\i}a P\'erez, A.Gonz\'alez-Arroyo, C.Pena
and~P.van Baal,~Phys.~Rev.~D60 (1999) 031901;~M.N.Chernodub,~T.C.Kraan
and P. van Baal, these proceedings, \verb+hep-lat/+ \verb+9907001+.

\bibitem{tonyrev}
A. Gonz\'alez-Arroyo, in {\em Advanced School for Nonperturbative Quantum Field
Physics}, eds. M. Asorey and A. Dobado (1998) p.57.

\bibitem{tonylat99}
M.~Garc\'{\i}a~P\'erez,~A.~Gonz\'alez-Arroyo, A.~Montero,
C.~Pena and P.~van~Baal, these proceedings.

\bibitem{NahmTBC}
A.~ Gonz\'alez-Arroyo, Nucl. Phys. B548 (1999) 626.

\bibitem{impcool}
M.~Garc\'{\i}a P\'erez, A.~ Gonz\'alez-Arroyo, J.~Snippe and P.~van~Baal,
Nucl. Phys. B413 (1994) 535.

\bibitem{nahm1}
A.~Gonz\'alez-Arroyo and C.~Pena, JHEP 09 (1998) 013.

\bibitem{BiCGStab}
H.~van der Vorst,
 SIAM J. Sc. Stat. Comp. 13 (1992) 631;
A.~Frommer, V.~Hanemann, B.~N\"ockel, Th.~Lippert and K.~Schilling,
 Int. J. Mod. Phys. C5 (1994) 1073.

\bibitem{dualities}
M.~Garc\'{\i}a P\'erez, A. Gonz\'alez-Arroyo,
C.~Pena and P.~van~Baal, \verb+hep-th/+ \verb+9905138+.

\bibitem{vortex}
A.~Gonz\'alez-Arroyo and A.~Montero, Phys. Lett. B442 (1998) 273;
A.~Montero, \verb+hep-lat/+ \verb+9906010+.


\end{thebibliography}
\end{document}